\documentclass[conference]{IEEEtran}
\IEEEoverridecommandlockouts
\usepackage{cite}
\usepackage{amsmath,amssymb,amsfonts}
\usepackage{algorithmic}
\usepackage{algorithm}
\usepackage{graphicx}
\usepackage{booktabs}
\usepackage{amsmath}
\usepackage{textcomp}
\usepackage{braket}
\usepackage{xcolor}
\usepackage{subcaption}
\def\BibTeX{{\rm B\kern-.05em{\sc i\kern-.025em b}\kern-.08em
    T\kern-.1667em\lower.7ex\hbox{E}\kern-.125emX}}
\begin{document}

\title{A Quantum Framework for K Coloring of Graphs\\
}

\author{\IEEEauthorblockN{Lord Sen}
\IEEEauthorblockA{\textit{Computer Science and Engineering} \\
\textit{National Institute of Technology Rourkela}\\
Rourkela, Odisha \\
123cs0226@nitrkl.ac.in}
\and
\IEEEauthorblockN{Shyamapada Mukherjee}
\IEEEauthorblockA{\textit{Computer Science and Engineering} \\
\textit{National Institute of Technology Rourkela}\\
Rourkela, Odisha \\
mukherjees@nitrkl.ac.in}
}

\maketitle

\begin{abstract}
Graph coloring is a well-known NP-complete problem with applications in scheduling, register allocation, frequency assignment, and network optimization. Quantum computing offers the potential for polynomial or even super-polynomial speedups in certain problem instances, yet practical quantum graph coloring methods must carefully balance qubit resources, circuit depth, and constraint enforcement. We proposed a solver agnostic quantum framework for $K$-coloring. Along with a novel encoding and efficient constraint implementation for its exact coloring exploiting Grover search and almost optimal coloring by the Quantum Approximate Optimization Algorithm \emph{(QAOA)}, and \emph{Quantum Annealing}.  Unlike using $O(NK)$ qubits as in \emph{SOTA}, we reduced the qubit requirements to $O(N \log_2 K)$, along with optimized comparator circuits enforcing adjacency constraints. Further symmetry-based graph reduction is incorporated as an optional preprocessing step to further reduce the instance size before quantum execution. 
\end{abstract}

\section{Introduction}
\label{sec:introduction}
Graph node coloring is the task of assigning one of $K$ distinct colors to each vertex of a graph $G = (V, E)$ such that no two adjacent vertices share the same color. The minimum number of colors required, known as the chromatic number $\chi(G)$, is central to complexity theory and combinatorial optimization. Since determining $\chi(G)$ is NP-complete, classical algorithms rely on heuristics, branch-and-bound, or exponential-time exact search for large graphs. Quantum computing offers new paradigms for solving combinatorial problems, including amplitude amplification~\cite{b1,b5} adiabatic optimization~\cite{b27}, and hybrid variational methods such as QAOA~\cite{b26}. These methods can, in principle, explore large search spaces more efficiently than classical algorithms. However, practical adoption for graph coloring faces two key challenges:
\begin{enumerate}
    \item \textbf{Qubit efficiency:} Conventional approaches and \emph{SOTA} requires $N \cdot K$ qubits, which becomes prohibitive for large graphs or color sets.
    \item \textbf{Constraint enforcement:} Quantum states must be restricted to valid colorings or penalized for violations, requiring efficient multi-qubit gates or cost Hamiltonians.
\end{enumerate}
To address these challenges we propose using $\log_2 K$ qu bits for each color, and for $N$ nodes, it makes a total of $N\cdot \log_2K$ qubits. But this has a problem, which is why $SOTA$ approaches tend to avoid this. This increases the ancilla qubits required for constraint implementation and makes the constraint implementation difficult, also adds additional constraints.
The main contributions are:
\begin{itemize}
    \item[1.] A solver-agnostic K-graph coloring formulation with 3 efficient solver specific architectures exploiting Grover search, QAOA, and quantum annealing.
    \item[2.] Optimized solver specific constraint implementer circuits reducing ancilla qubits.
    \item[3.] Efficient encoding of colors reducing logical qubit count to $O(N \log_2 K)$ while maintaining compatibility with multiple quantum solvers and also reducing the ancilla qubits.
    \item[4.] An optional classical pre-processing step exploiting symmetries to reduce problem size, qubit count and solution space.
\end{itemize}
\section{Related Work}
\label{sec:related}
Quantum algorithms for graph coloring have been explored in various contexts, primarily focusing on specific graph classes or encoding schemes \cite{b17, b18}, while subsequent studies extended these ideas to more general graphs~\cite{b1}, \cite{b19, b20, b21, b261, b23, b24, b25, b3}.
Recent approaches have focused on hybrid quantum-classical methods, such as the Quantum Approximate Optimization Algorithm (QAOA)~\cite{b26}, which has been applied to graph coloring problems~\cite{b2}. These methods typically use one-hot encoding of colors, leading to high qubit requirements.
Other works have explored quantum annealing for graph coloring, leveraging the Ising formulation of the problem~\cite{b6}, \cite{b9}. Symmetric techniques have been explored in~\cite{sencol, b10, b11}. However, these methods often overlook qubit efficiency and effective constraint enforcement, leading to a low valid fraction and requiring traversal through many invalid colorings. In contrast, our framework significantly reduces qubit usage while maintaining compatibility with multiple quantum solvers and reducing ancillas with optimized comparator circuits for enforcing constraints, making the method more practical for near-term quantum devices with higher valid fraction. This paper is an extension of our paper~\cite{senquant}.
\section{Proposed Quantum Graph Coloring Framework}
\label{sec:proposed}

We propose a quantum algorithm for $K$-coloring of an undirected graph $G = (V, E)$ that emphasizes efficient quantum encoding of color assignments and compatibility with multiple quantum solvers. The input consists of an undirected graph $G = (V, E)$ with $N = |V|$ vertices and $M = |E|$ edges, and a target number of colors $K \geq \chi(G)$, where $\chi(G)$ is the chromatic number of $G$. The quantum resource constraints include available qubits and circuit depth. The proposed pipeline as in Algorithm~\ref{alg:hybrid-coloring} comprises five stages:
\begin{algorithm}[htbp]
\caption{Hybrid Classical - Quantum Graph Coloring}
\label{alg:hybrid-coloring}
\begin{algorithmic}[1]
\REQUIRE $G=(V,E)$, $N\leftarrow |V|$, $M\leftarrow |E|$, $K\ge \chi(G)$
\STATE $G_q(V_q,E_q)=Reduce(G)$, $N_q\leftarrow |V_q|$.
\STATE $Q_{logical}\leftarrow Encode(G_q)$
\FORALL{$(u,v)\in E_q$}
\STATE $Q_{ancilla}\leftarrow Constraint(G_q)$
\ENDFOR
\STATE $S_{col}\leftarrow Solver(G_q, Q_{\mathrm{ancilla}}, Q_{\mathrm{logical}})$
\STATE $I_{valid}, S_{final}\leftarrow PostProcess(S_{col})$
\IF{$I_{valid}$} 
\RETURN $S_{final}$
\ENDIF
\end{algorithmic}
\end{algorithm}
\begin{algorithm}[]
\caption{Encoding \& Constraint Implementation}
\label{alg:constraint}
\begin{algorithmic}[1]
\REQUIRE $G_q=(V_q,E_q)$, $N_q\leftarrow |V_q|$, $K\ge \chi(G_q)$,  $Q_{\max}\leftarrow$ Available qubits, $A_Q\leftarrow$ Quantum Solver (one of $Grover_{FQ}$: Full-Quantum Grover, $Grover_{QC}$: Grover with Quantum-equality/Classical-range constraints, $QAOA$, or $QA$: Quantum Annealing; see Section~\ref{sec:solverspec})
\IF{$Q_{\max} \ge N_q\cdot K$ \AND $A_Q\in \{Grover_{FQ}, QA\}$}
\STATE $Q_{\mathrm{logical}} \leftarrow N_q \cdot K$ qubits.
\ELSE
\STATE $Q_{\mathrm{logical}} \leftarrow N_q \cdot \lceil\log_2 K\rceil$ qubits. 
\ENDIF
\FORALL{$(u,v)\in E_q$}
\IF{$Q_{\mathrm{logical}} ==N_q \cdot K$}
\STATE $Q_{ancilla}\leftarrow Constraint_{one}(G_q)$
\ELSE
\STATE $Q_{ancilla}\leftarrow Constraint_{bin}(G_q)$
\ENDIF
\ENDFOR
\end{algorithmic}
\end{algorithm}

\subsection{Graph Reduction}
A symmetry detection algorithm may be applied to identify vertex orbits as in \cite{b4}, \cite{b13}, \cite{b14}. Vertices in the same orbit are collapsed into a single representative, yielding a quotient graph $G_q$. This step reduces the number of vertices $N$ before quantum encoding. We have presented a few such methods in Section \ref{sec:Red_methods}, though these are not essential to our approach but can increase performance significantly. The quantum algorithm operates on $G_q$, but the final solution is lifted back to $G$ by assigning the same color to all vertices in each orbit.

\subsection{Quantum Encoding of Colors}
\label{sec:binaryEnc}
Let $N = |V|$. We consider two primary encoding schemes, the first one is used in most \emph{SOTA} models called One-Hot Encoding. Here, each vertex $v_i$ is assigned $K$ qubits $\{q_{i,1}, \dots, q_{i,K}\}$, one for each color. Exactly one qubit is in the $\ket{1}$ state, enforcing a single color choice $ \sum_{j=1}^K q_{i,j} = 1, \quad \forall i \in \{1, \dots, N\}$.
The total qubit requirement is $Q_{\text{1hot}} = N \cdot K.$

We propose \emph{Binary Encoding}, where each vertex $v_i$ uses $b=\lceil \log_2 K \rceil$ qubits to represent a color index in binary form. The total qubit requirement is:
\begin{equation}
Q_{\text{bin}} = N \cdot \lceil \log_2 K \rceil=N\cdot b.
\end{equation}
Binary encoding reduces qubits significantly for larger $K$, but requires additional comparator circuits to enforce coloring constraints. To binary encode, we define a binary register $b_{i,1}, \dots, b_{i,\lceil \log_2 K \rceil}$ for each vertex $v_i$, where the value of the register corresponds to the color index.  Out-of-range values (i.e., $\ge K$) are invalid and penalized in the cost function.

\subsection{Constraint Implementation}
\label{sec:constraint}
Let $c(u)$ denote the color of vertex $u$. For each edge $(u,v) \in E$, the penalty term, called \emph{Edge Equality flag} as,
\begin{equation}
P_{uv} =
\begin{cases}
1, & \text{if } c(u) = c(v), \\
0, & \text{otherwise}.
\end{cases}
\end{equation}
In one-hot encoding, $P_{uv}$ is implemented via multi-controlled Toffoli gates detecting matching active color qubits. In binary encoding, comparators check equality between binary color registers. Efficient constraint implementation is at the heart of practical quantum graph coloring, since naive multi-controlled implementations of $P_{uv}$ scale poorly with $b$ and $M$. Solver-specific, efficient constraint-implementation circuits are presented in Section~\ref{sec:solverspec}.

\subsection{Quantum Solvers}

\subsubsection{Grover Search with Constraint Oracle}
The search space comprises all possible color assignments, and the oracle marks states satisfying all constraints $P_{uv} = 0$. Amplitude amplification yields a valid coloring in $O(\sqrt{P})$ iterations, where $P$ is the size of the feasible search space.

\subsubsection{QAOA with Coloring Hamiltonian}
The cost Hamiltonian is,
$H_{\text{cost}} = \sum_{(u,v) \in E} \lambda \cdot P_{uv},$ where $\lambda > 0$ is a penalty strength. Mixers are chosen to preserve feasibility - XY-mixers for one-hot encoding, and constraint-preserving binary mixers for binary encoding.

\begin{figure*}[htbp]
    \centering
    \includegraphics[width=0.82\linewidth, height=5cm]{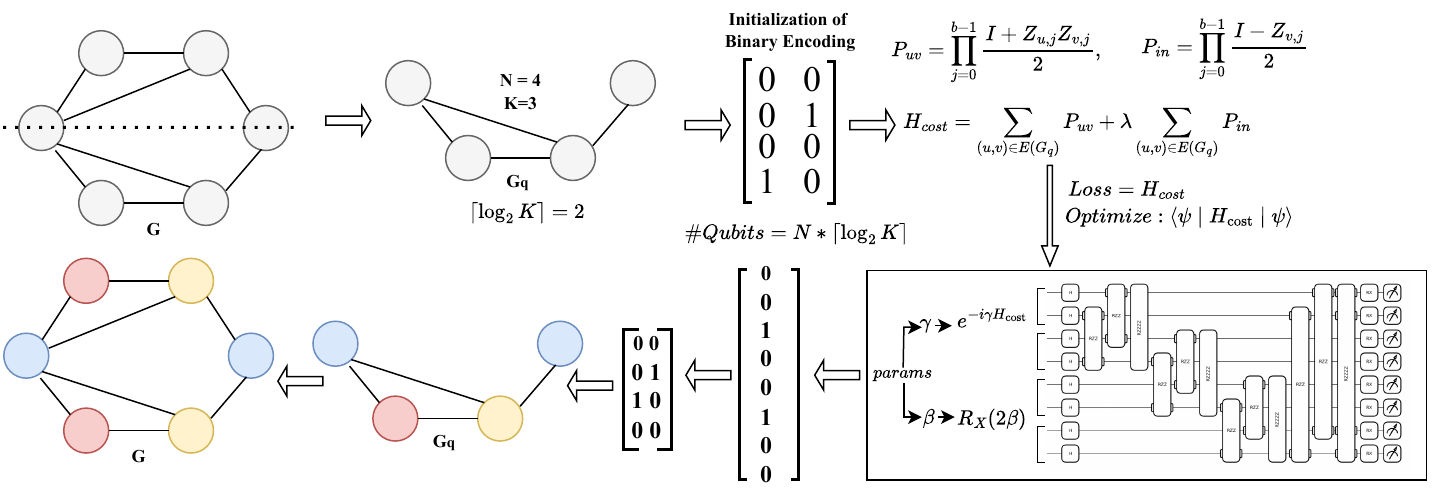}
    \caption{Proposed Quantum Graph Coloring framework. The framework is solver agnostic but the shown \emph{QC} is of \emph{QAOA}.}
    \label{fig:method}
\end{figure*}

\subsubsection{Quantum Annealing}
An equivalent Ising formulation is constructed using the same $H_{\text{cost}}$, with penalty strengths ensuring no adjacent vertices share the same color.\\
Each of these methods have been explained in detail in Section~\ref{sec:solverspec}.

\subsection{Post-Processing}
All qubits are measured to obtain a candidate coloring. In the case of quotient graph preprocessing, the solution is lifted back to $G$ by assigning the same color to vertices in the same orbit as shown in Fig.~\ref{fig:method}. Approximate solvers such as QAOA may require a classical repair step (e.g., greedy recoloring) to operate in lesser iterations.

\section{Solver Specific Architectures}
\label{sec:solverspec}
In this section, we present solver-specific circuit architectures for Grover search, QAOA, and quantum annealing, and analyze which combination of solver and encoding is best suited to which graph regime (summarized in Table~\ref{tab:compare} and Algorithm~\ref{alg:solver}).
\subsection{Grover-Based Quantum Architecture}

The circuit encodes vertex colors into qubits, detects constraint violations using ancillas, and applies amplitude amplification to highlight valid colorings.

\subsubsection{Encoding}
Consider $G = (V,E)$ with $|V|$ vertices and $|E|$ edges. For $K$ colors, each vertex requires $b = \lceil \log_2 K \rceil$ qubits to represent a color.  Thus, the logical register uses $n_{\text{logical}} = N \times b$ qubits. 

\subsubsection{Ancilla Allocation}
The circuit uses additional ancilla qubits for constraint checking; each \emph{edge equality flag} (Section~\ref{sec:constraint}) requires $b$ XOR ancillas and one flag qubit per edge. For \emph{vertex invalidity flags}, $\forall$ vertex $v$, a multi-controlled NOT marks the state $c(v)=K$, requiring one flag qubit per vertex. Local flags are combined into a single global ancilla, flipped only if all constraints are satisfied. \emph{Work ancillas} are additionally used to implement the multi-controlled gates efficiently (via Toffoli decomposition).

Two allocation regimes are possible, trading circuit width for depth:
\begin{itemize}
    \item \textbf{Sequential (compute–uncompute):} edge and vertex flags are computed one at a time and immediately uncomputed after being folded into the global ancilla. This keeps the \emph{peak} ancilla width at $n_{\text{ancilla}}^{\text{peak}} = O(b)$ (one edge's worth of XOR ancillas plus $O(1)$ flag/work qubits, reused across all $M$ edges and $N$ vertices), at the cost of $O(M+N)$ sequential compute/uncompute blocks and hence added circuit depth.
    \item \textbf{Parallel:} all edge and vertex flags are computed simultaneously to minimize depth, requiring $n_{\text{ancilla}} = O(Mb+N)$ ancillas held live at once, matching the depth bound $O(Mb+Nb)$ reported in Table~\ref{tab:compare}.
\end{itemize}
Hence the total qubit count is $n_{\text{total}} = n_{\text{logical}} + n_{\text{ancilla}}$, where $n_{\text{ancilla}}$ ranges from $O(b)$ (width-optimized, sequential) to $O(Mb+N)$ (depth-optimized, parallel) depending on which resource -- qubits or circuit depth -- is the binding constraint on the target device. This distinction is what Algorithm~\ref{alg:solver} checks via $Q_{\max}$ and $D_{\max}$.

\subsubsection{Oracle Construction}
The oracle is responsible for marking valid colorings.  
For each edge $(u,v)$, we compute $f_{uv}$ and $g_v$ for each vertex $v$,
\begin{equation}
    f_{uv} = \begin{cases}
        1 & \text{if } c(u)=c(v), \\
        0 & \text{otherwise},
    \end{cases} ,\quad
    g_v = \begin{cases}
        1 & \text{if } c(v) \geq K, \\
        0 & \text{otherwise}.
    \end{cases}
\end{equation}
The global flag is then given by
\begin{equation}
    F = \prod_{(u,v)\in E} (1-f_{uv}) \cdot \prod_{v\in V} (1-g_v).
\end{equation}
When $F=1$, the assignment is a valid $K$-coloring, in this case the oracle applies a phase flip:
\begin{equation}
    \mathcal{O} \ket{x} = (-1)^{F(x)} \ket{x}.
\end{equation}


Depending on how constraints are enforced, two main strategies arise:

\paragraph{Full Quantum Constraint Enforcement ($Grover_{FQ}$)}
Both vertex-color validity and edge inequality constraints are implemented within the quantum oracle. Here, the ancilla overhead grows significantly with graph size and $K$ as $O(N\cdot b+E)$.

\paragraph{Quantum Equality Checks with Classical Range Validation ($Grover_{QC}$)}
A hybrid approach reduces quantum overhead where discarding colors $\geq K$ is moved to post-processing. Only edge constraints are implemented in the quantum circuit. This reduces the number of ancillas by $O(N\cdot b)$ and circuit complexity compared to the full-quantum method, while still leveraging quantum superposition for enforcing adjacency constraints. We use $Grover_{FQ}$ and $Grover_{QC}$ throughout to refer to these two strategies unambiguously.

\subsubsection{Diffusion Operator}
The diffusion operator performs an inversion about the mean amplitude. It consists of first applying Hadamard gates to all logical qubits, then Flipping all qubits with Pauli-$X$ and applying a multi-controlled $Z$ gate on the all-zero state:$D = H^{\otimes n} \, X^{\otimes n} \, (I - 2\ket{0}\bra{0}) \, X^{\otimes n} \, H^{\otimes n}$.

\begin{figure}[htbp]
    \centering
    \includegraphics[width=0.9\linewidth, height=5cm]{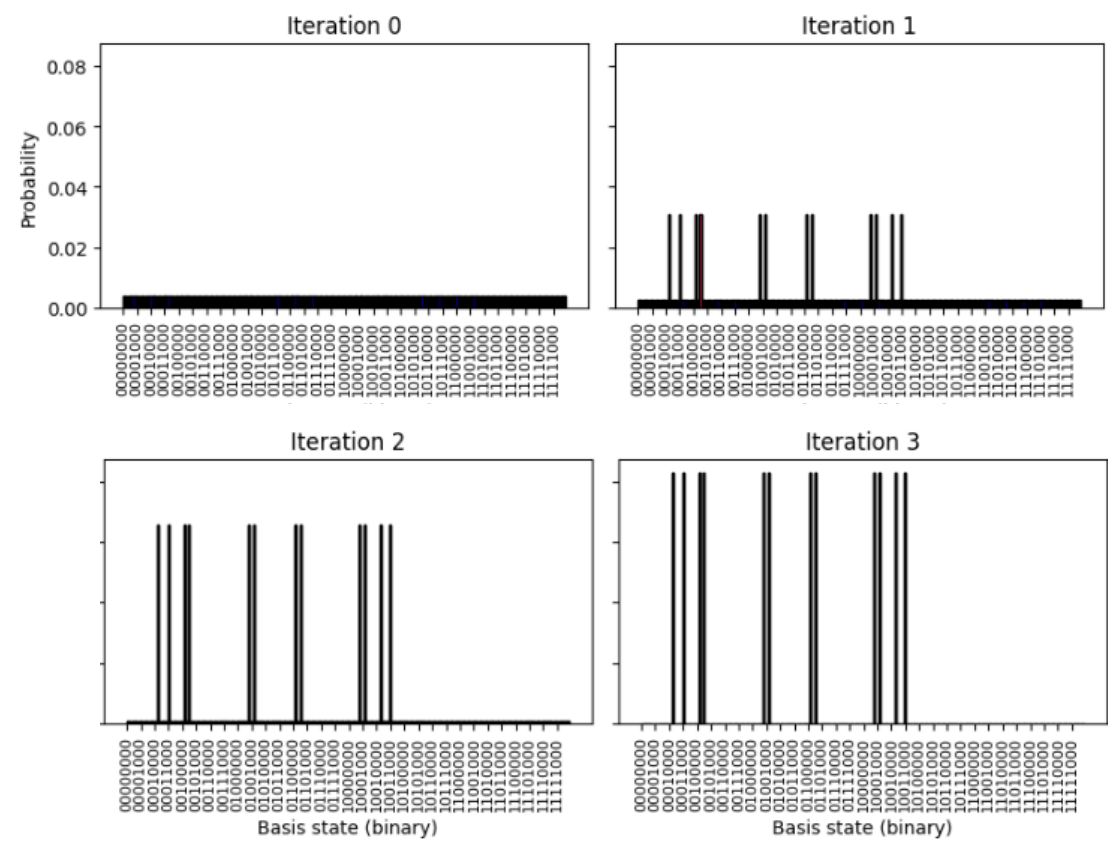}
    \caption{Amplification of valid states over successive Grover iterations. The $y$-axis shows the measurement probability of each basis state and the $x$-axis enumerates candidate colorings (binary basis states); only every 8th basis state is labeled to avoid tick-label congestion.}
    \label{fig:grover_amp}
\end{figure}

Each Grover iteration applies the oracle followed by the diffusion operator as shown in Fig.~\ref{fig:grover_amp}. In our implementation, we performed 3 Grover iterations on a $25$-qubit simulator with $2000$ shots. Experimental results showed that valid solutions were consistently amplified, and their frequency among measurement outcomes was significantly higher compared to invalid configurations as in Fig.~\ref{fig:grover_amp}.

\subsection{QAOA Implementation for Graph Coloring}
\label{sec:qaoa}

QAOA provides a variational framework to approximate solutions to combinatorial problems. In general, the method alternates between applying a \emph{cost unitary}, which encodes the problem constraints, and a \emph{mixer unitary}, which drives exploration of the solution space to minimize the expected cost Hamiltonian. This subsection presents an efficient QAOA formulation for $K$-coloring that minimizes the logical qubit count, reduces the ancilla overhead for constraint implementation to $O(1)$, and empirically converges to a valid coloring in relatively few layers. As discussed in Section~\ref{sec:proposed}, we begin with the encoding of the colors to be assigned.

\subsubsection{Encoding}
We adopt binary encoding for efficiency, as discussed in Section~\ref{sec:binaryEnc}. However, binary encoding raises the risk of increased ancilla overhead when equality constraints are enforced via comparator circuits (as in the Grover oracle of Section~\ref{sec:solverspec}A). The following subsections show how QAOA's Hamiltonian formulation sidesteps this cost, reducing the ancilla requirement for constraint implementation to $O(1)$.

\subsubsection{Cost Hamiltonian}
For each edge $(u,v)\in E$, a penalty is applied if both vertices receive the same color. In binary encoding, equality of two registers is enforced by bitwise comparators as in $P_{uv}$ and range of colors by $P_{in}$:
\begin{equation}
P_{uv} = \prod_{j=0}^{b-1}\frac{I + Z_{u,j}Z_{v,j}}{2},\quad
P_{\mathrm{in}}(v)
\;=\;
\prod_{j=0}^{b-1}
\frac{I - Z_{v,j}}{2}
\end{equation}

The total cost Hamiltonian is therefore

\begin{equation}
H_{cost} \;=\; \sum_{(u,v)\in E} P_{\mathrm{uv}}(u,v) \;+\; \lambda \sum_{v} P_{\mathrm{in}}(v)
\end{equation}
where $\lambda>0$ is a tunable penalty weight. This operator assigns energy penalties to invalid colorings while leaving valid assignments in the ground state manifold.

\subsubsection{QAOA Layers}
The depth-$p$ QAOA circuit alternates between the following unitaries as shown in Fig.~\ref{fig:method}:
\begin{align}
U_C(\gamma_\ell) &= e^{-i\gamma_\ell H_{\text{cost}}}, \quad
U_M(\beta_\ell) = \prod_{q=1}^{Nb} R_X(2\beta_\ell),
\end{align}
where $U_C$ encodes the coloring constraints and $U_M$ mixes amplitudes by rotating each qubit around the $X$-axis. The circuit begins from a uniform superposition state $\ket{\psi_0} = \bigotimes_{q=1}^{Nb} H\ket{0}$,
and evolves to get the valid colorings as
\begin{equation}
\ket{\psi_p} = \left[\prod_{\ell=1}^{p} U_M(\beta_\ell) U_C(\gamma_\ell)\right]\ket{\psi_0}.
\end{equation}

\subsubsection{Optimization and Decoding}
Classical optimization updates $(\gamma,\beta)$ to minimize $\langle H_{\text{cost}}\rangle$. Measurement yields a bitstring of length $Nb$, which is decoded into vertex colors as in Fig.~\ref{fig:method}. Invalid assignments are discarded or penalized. The most frequently observed valid coloring corresponds to the approximate solution.

QAOA thus constructs a quantum state whose amplitude distribution is biased toward valid colorings of the input graph. Low-depth instances ($p=1,2$) already provide nontrivial performance improvements in small simulations. The choice of binary versus one-hot encoding balances qubit count and circuit depth, making QAOA a versatile approach for $K$-coloring in near-term quantum devices.

\subsection{Quantum Annealing (QA) for Graph Coloring}
\label{sec:qa-coloring}

\subsubsection{Ising Formulation}
Given a graph $G=(V,E)$ and a target number of colors $K$, we encode the $K$-coloring problem as a quadratic unconstrained binary optimization (QUBO) using one-hot encoding. For each vertex $v\in V$ and color $k\in\{0,\ldots,K-1\}$, we introduce a binary variable $x_{v,k}\in\{0,1\}$ that indicates whether vertex $v$ is assigned color $k$. Now, each vertex must choose exactly one color:
\begin{equation}
\sum_{k=0}^{K-1} x_{v,k} = 1
\quad\Rightarrow\quad
H_{\text{vert}}(v) \;=\; \lambda_v\Big(\sum_{k} x_{v,k}-1\Big)^2,
\label{eq:vertex}
\end{equation}
and adjacent vertices must not share the same color,
\begin{equation}
H_{\text{edge}}(u,v) \;=\; \lambda_e \sum_{k=0}^{K-1} x_{u,k}x_{v,k},\quad (u,v)\in E.
\label{eq:edge}
\end{equation}

Hence, the total Ising energy, that annealer minimizes is
\begin{equation}
H(x) \;=\; \sum_{v\in V} H_{\text{vert}}(v) \;+\; \sum_{(u,v)\in E} H_{\text{edge}}(u,v),
\label{eq:totalQubo}
\end{equation}
with penalty weights $\lambda_v,\lambda_e>0$. Expanding~\eqref{eq:vertex} produces linear and quadratic terms only. Thus the model is compatible with Ising hardware without auxiliary variables. The resulting QUBO can be mapped to an Ising Hamiltonian
\(
H(s)=\sum_i h_i s_i + \sum_{i<j} J_{ij}s_is_j
\)
via $x_i=(1+s_i)/2$ where $h_i$ is the effective local field on spin $i$, $J_{ij}$ is effective interaction strength between spins $i$ and $j$ with spin variables $s_i$ and $s_j$.

\subsubsection{Annealing Schedule}
Given an Ising problem $(h,J)$, a quantum annealer implements the time-dependent Hamiltonian
$$H(t) = A(t)\sum_i X_i + B(t)\big(\sum_i h_i Z_i + \sum_{i<j} J_{ij} Z_i Z_j\big),$$
interpolating from a transverse-field driver to the problem Hamiltonian. Here, $X_i,Z_i$ are Pauli-X and Z operators and $A(t), B(t)$ are schedules controlling transverse field and problem Hamiltonian. The anneal time $T$, number of reads can be tuned to balance solution quality and throughput. $\lambda_v$ and $\lambda_e$ are chosen to get optimal performance. While, binary encoding reduces qubits to $N\lceil\log_2K\rceil$, but equality tests between color indices are higher-degree polynomials.
\section{Reduction methods}
\label{sec:Red_methods}

\begin{figure*}[htbp]
    \centering
    \includegraphics[width=0.8\linewidth, height=6cm]{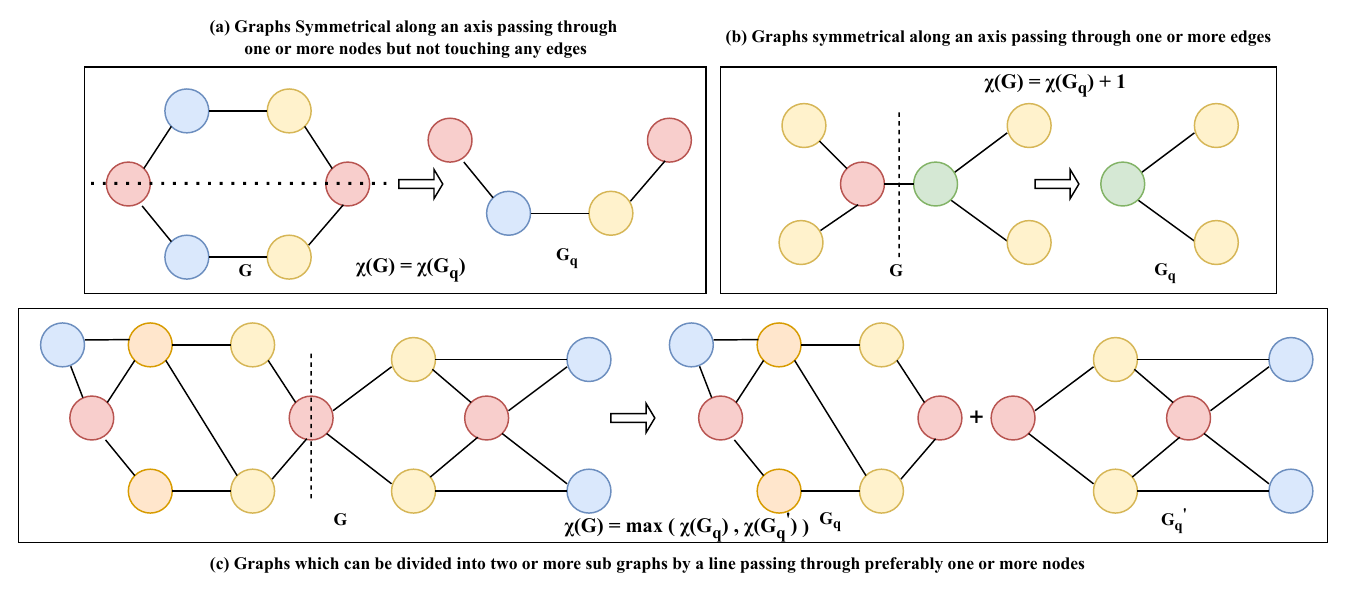}
    \caption{Different reduction methods and special graphs that can be used to simplify the problem. These graphs can help in reducing the number of nodes and qubits required for the quantum circuit. We call them Type-1, Type-2, and Type-3 graphs.}
    \label{fig:Redn}
\end{figure*}

In this section, we discuss classical pre-processing reductions that exploit graph symmetry to shrink $G$ before quantum encoding, reducing qubit count, circuit depth, and solution-space size without changing the chromatic number. Throughout, $G=(V,E)$ denotes the original graph with $|V|=N$, and a \emph{symmetric axis} (or \emph{symmetric line}) refers to an automorphism-induced reflection of $G$ that partitions $V$ into mirrored pairs, together with any nodes or edges lying exactly on the axis. We consider three structural cases, illustrated in Fig.~\ref{fig:Redn}:
\begin{itemize}
    \item \textbf{Type-1 graph:} the symmetric axis passes through one or more \emph{nodes} of $G$ but through no edges (Fig.~\ref{fig:Redn}a). Folding $G$ along this axis identifies each mirrored pair of vertices, producing a single \emph{daughter graph} (quotient graph) $G_q$ with $\chi(G)=\chi(G_q)$.
    \item \textbf{Type-2 graph:} the symmetric axis passes through one or more \emph{edges} of $G$ and through no nodes (Fig.~\ref{fig:Redn}b). Here the axis bisects the graph into two isomorphic halves without a shared vertex; folding introduces one new edge per bisected edge, so in the worst case $\chi(G)=\chi(G_q)+1$ (equality with $\chi(G)=\chi(G_q)$ holds whenever the folded edge's endpoints are not already forced to the same color in an optimal coloring of $G_q$).
    \item \textbf{Type-3 graph:} $G$ can be split into two vertex-induced subgraphs $G_q=(V_q,E_q)$ and $G_q'=(V_q',E_q')$ joined at a single cut vertex (Fig.~\ref{fig:Redn}c), so $\chi(G) = \max\big(\chi(G_q),\chi(G_q')\big)$, since a valid coloring of each piece can always be aligned at the shared vertex by permuting one piece's color labels.
\end{itemize}
Each case yields a formula, stated below, for how many qubits are saved when the corresponding symmetry is present. As with any symmetry-exploiting preprocessing, finding the relevant automorphisms is itself a graph-automorphism / canonical-labeling problem; this is not known to run in polynomial time for general graphs in the worst case, though it is efficient in practice and for structured instances (e.g., using \texttt{nauty}-style canonical labeling~\cite{b13}). We therefore treat this reduction as an optional, best-effort preprocessing step (Algorithm~\ref{alg:hybrid-coloring}, line 1) rather than a guaranteed one: when no exploitable symmetry is found, the pipeline proceeds directly on $G$ ($G_q=G$).

\medskip
\textbf{Result 1 (Type-1 reduction).} \emph{If, for a Type-1 graph $G=(V,E)$ with $|V|=N$ and $K$ colors, the symmetric axis passes through $m$ nodes, then:}
\begin{enumerate}
    \item[(a)] \emph{The daughter graph $G_q$ has $\frac{N+m}{2}$ nodes.}
    \item[(b)] \emph{$N$ and $m$ have the same parity (both even or both odd).}
    \item[(c)] \emph{The qubit count for binary encoding decreases by $\frac{N-m}{2}\cdot\lceil\log_2 K\rceil$.}
\end{enumerate}
\emph{Proof sketch.} The axis fixes $m$ nodes and pairs the remaining $N-m$ nodes into $\frac{N-m}{2}$ mirrored pairs, each collapsing to a single representative under the automorphism. Hence $|V_q| = m + \frac{N-m}{2} = \frac{N+m}{2}$, giving (a). Since $\frac{N-m}{2}\in\mathbb{Z}$, $N-m$ is even, so $N$ and $m$ share parity, giving (b). Binary encoding uses $\lceil\log_2 K\rceil$ qubits per vertex (Section~\ref{sec:binaryEnc}), so the qubit count drops from $N\lceil\log_2 K\rceil$ to $\frac{N+m}{2}\lceil\log_2 K\rceil$, a reduction of $\big(N-\frac{N+m}{2}\big)\lceil\log_2 K\rceil = \frac{N-m}{2}\lceil\log_2 K\rceil$, giving (c). $\blacksquare$

\medskip
\textbf{Result 2 (Type-2 reduction).} \emph{If, for a Type-2 graph $G=(V,E)$ with $|V|=N$ and $K$ colors, the symmetric axis passes through $m$ edges (and no nodes), then:}
\begin{enumerate}
    \item[(a)] \emph{$N$ is even, and the daughter graph has $\frac{N}{2}$ nodes, independent of $m$.}
    \item[(b)] \emph{The qubit count for binary encoding decreases by $\frac{N}{2}\cdot\lceil\log_2 K\rceil$.}
\end{enumerate}
\emph{Proof sketch.} Since the axis touches no node, every vertex is paired with a distinct mirror vertex, so $N$ must be even and folding halves the vertex count to $\frac{N}{2}$ regardless of how many edges ($m$) cross the axis, giving (a); $m$ only affects how many new edges appear in $G_q$ (and hence whether $\chi(G)=\chi(G_q)$ or $\chi(G_q)+1$, per the Type-2 definition above), not the vertex count. The qubit saving follows exactly as in Result~1(c) with the node count on the axis set to $0$: reduction $=\big(N-\frac{N}{2}\big)\lceil\log_2K\rceil=\frac{N}{2}\lceil\log_2K\rceil$, giving (b). $\blacksquare$

\medskip
\textbf{Result 3 (mixed node/edge axis).} \emph{If the symmetric axis of $G=(V,E)$, $|V|=N$, passes through $m$ nodes and $p$ edges simultaneously, then:}
\begin{enumerate}
    \item[(a)] \emph{The daughter graph has $\frac{N+m}{2}$ nodes, independent of $p$.}
    \item[(b)] \emph{$N$ and $m$ have the same parity.}
    \item[(c)] \emph{The qubit count for binary encoding decreases by $\frac{N-m}{2}\cdot\lceil\log_2 K\rceil$.}
\end{enumerate}
\emph{Proof sketch.} This is the general case combining Results 1 and 2: the $m$ nodes on the axis behave exactly as in Result~1, fixing $\frac{N+m}{2}$ as the daughter vertex count, while the $p$ edges crossing the axis (as in Result~2) affect only the daughter graph's \emph{edge} set and possibly its chromatic number, not its vertex count. The vertex-count and qubit-count formulas are therefore identical to Result~1, with $p$ playing no role, which is why (a)--(c) do not depend on $p$. $\blacksquare$

\medskip
\textbf{Result 4 (Type-3 reduction).} \emph{If a Type-3 graph $G$ decomposes into vertex-induced subgraphs $G_q=(V_q,E_q)$ and $G_q'=(V_q',E_q')$ sharing exactly one cut vertex, then encoding $G_q$ and $G_q'$ separately (solving each on the quantum device independently, with the shared vertex's color fixed once and reused) reduces the qubit count by $\max(|V_q|,|V_q'|)\cdot\lceil\log_2 K\rceil$ relative to encoding $G$ as a single instance.}

\emph{Proof sketch.} Since $\chi(G)=\max(\chi(G_q),\chi(G_q'))$ (established above), a valid $K$-coloring of $G$ is obtained by independently coloring $G_q$ and $G_q'$ and then permuting the color labels of one piece so the shared cut vertex agrees in both. Because $G_q$ and $G_q'$ share only that one vertex, the harder of the two subproblems -- whichever has more vertices, $\max(|V_q|,|V_q'|)$ -- determines the coloring difficulty of $G$, while the smaller piece can be handled largely independently (even classically, if small enough). Encoding only the larger subgraph on the quantum device needs $\max(|V_q|,|V_q'|)\cdot\lceil\log_2K\rceil$ qubits, versus $(|V_q|+|V_q'|-1)\cdot\lceil\log_2K\rceil = (N-1)\lceil\log_2 K\rceil$ qubits for encoding $G$ directly (the $-1$ accounts for the shared cut vertex being counted once), yielding the stated saving. $\blacksquare$

\medskip
Results 1--4 assume a single reduction pass; each can be applied recursively to $G_q$ (or to $G_q$ and $G_q'$) whenever further symmetry is present, compounding the qubit savings, as quantified for a concrete example in the Reduction Analysis of Section~\ref{sec:analysis}.

\begin{algorithm}[t]
\caption{Quantum Solver Selection for $K$-Coloring}
\label{alg:solver}
\begin{algorithmic}[1]
\REQUIRE $G=(V,E)$, $N=|V|$, $M=|E|$, $\Delta_{\max}\gets$ max degree, $K\geq \chi(G)$, $Q_{\max}\gets$ available qubits, $D_{\max}\gets$ max circuit depth, $r\gets$ number of valid $K$-colorings of $G$ (feasible search-space size), $p\gets$ QAOA depth, $D\gets$ per-round circuit depth, $Q_{\text{anc}}\gets$ ancilla qubits required (Section~\ref{sec:solverspec})
\ENSURE $S_{\text{col}}$ (coloring solution)
\STATE $b \gets \lceil \log_2 K \rceil$, \; $Q_{\text{bin}} = N b$, \; $Q_{\text{1hot}} = N K$
\STATE $T_G \gets \tfrac{\pi}{4}\sqrt{2^{Nb}/r}$ \; \COMMENT{Grover iterations, Section~\ref{sec:solverspec}A}
\IF{$Q_{\text{bin}}+Q_{\text{anc}} \leq Q_{\max}$ \AND $T_G D \leq D_{\max}$ \AND $M \lesssim N \Delta_{\max}$}
    \STATE $S_{\text{col}} \gets \text{Grover}(G, Q_{\text{bin}}, Q_{\text{anc}})$ \hfill \COMMENT{sparse/moderate $M$}
\ELSIF{$Q_{\text{bin}} \leq Q_{\max}$ \AND $pD \leq D_{\max}$ \AND $\Delta_{\max} \ll N$}
    \STATE $S_{\text{col}} \gets \text{QAOA}(G, Q_{\text{bin}})$ \hfill \COMMENT{bounded-degree graphs}
\ELSIF{$Q_{\text{1hot}} \leq Q_{\max}$ \AND $N$ is moderate}
    \STATE $S_{\text{col}} \gets \text{QA}(G, Q_{\text{1hot}})$ \hfill \COMMENT{medium $N$, sparse $M$}
\ELSE
    \STATE $S_{\text{col}} \gets \emptyset$ \quad \COMMENT{not feasible with available resources}
\ENDIF
\end{algorithmic}
\end{algorithm}

Algorithm~\ref{alg:solver} should be applied to $G_q$ (the output of the reduction step, or $G$ itself if no reduction is found), substituting $N \to N_q = |V_q|$ throughout, so that the resource thresholds reflect the post-reduction problem size.
\section{Analysis}
\label{sec:analysis}
Let $N_{\text{shots}}$ denote the total number of measurement shots used in an experiment, and $N_{\text{valid}}$ the number of those shots corresponding to a valid coloring. The \emph{valid fraction} $\eta_v$ is then
\begin{equation}
\eta_v \;=\; \frac{N_{\text{valid}}}{N_{\text{shots}}}, \qquad \%\text{valid} = \eta_v \times 100.
\end{equation}

\begin{table*}[t]
\centering
\caption{Resource comparison across solvers and encodings. Here $G=(V,E)$ with $N=|V|$, $M=|E|$, $b=\lceil\log_2 K\rceil$, $\Delta_{\max}$ is the maximum vertex degree, and $r$ is the number of valid $K$-colorings of $G$ (the feasible search-space size used in the Grover iteration count). $Q_{\text{ancilla}}$ reports the \emph{peak} simultaneous ancilla count under depth-optimized (parallel) constraint evaluation (Section~\ref{sec:solverspec}); it can be reduced toward $O(b)$ at the cost of additional sequential depth via compute--uncompute reuse.}
\label{tab:compare}
\small
\begin{tabular}{@{}llcccccc@{}}
\toprule
Method & Encoding & $Q_{\text{logical}}$ & $Q_{\text{ancilla}}$ & Depth/round $D$ & Iter./Rounds & Runtime $\sim$ & Suitability \\
\midrule
Grover ($Grover_{FQ}$) & Binary & $N b$ & $O(Mb)$ & $O(Mb+Nb)$ & $T_G\!=\!\tfrac{\pi}{4}\sqrt{\tfrac{2^{Nb}}{r}}$ & $T_G(D_{\text{oracle}}+D_{\text{diff}})$ & small $N$, sparse $M$ \\
Grover ($Grover_{FQ}$) & One-hot & $N K$ & $O(M)$ & $O(MK+NK)$ & $T_G\!=\!\tfrac{\pi}{4}\sqrt{\tfrac{K^N}{r}}$ & $T_G(D_{\text{oracle}}+D_{\text{diff}})$ & small $N$, sparse $M$ \\
QAOA & Binary & $N b$ & $O(1)$ & $O(M2^b)$ & $p$ layers & $pD S_{\text{shots}}$ & Ideal \\
QAOA & One-hot & $N K$ & $O(1)$ & $O(MK+N K)$ & $p$ layers & $pD S_{\text{shots}}$ & Good \\
QA & One-hot & $N K$ & $0$ & embed-dep. & anneal time $T_a$ & $R T_a+T_{\text{embed}}$ & medium $N$ \\
QA & Binary & $Nb$+$Q_{\text{anc}}$ & $O(Mb)$ & embed-dep. & $T_a$ & $R T_a+T_{\text{embed}}$ & Not Ideal \\
\bottomrule
\end{tabular}
\end{table*}

\subsection{Resource Analysis}
Table~\ref{tab:compare} compares \# qubits and circuit complexity for the two encodings, across the three algorithms proposed. From the table, it is clear that the binary encoding is more qubit efficient, especially for large $K$ and \emph{QAOA} and reduces the circuit depth for Grover as in Table~\ref{tab:compare}. However, at the cost of increased circuit complexity for \emph{full quantum Grover's method} and \emph{QA}, where one-hot encoding, while requiring more qubits, offers simpler constraint implementation and is more suitable for small $K$ and \emph{QA}. The circuit depth of \emph{QAOA}, can be analyzed from Table~\ref{tab:resource}, with at least 4-5 layers \emph{QAOA} outputs valid colorings. However, with increasing layers $\eta_v$ increases, i.e, the likelihood of an output state to be valid increases till it overfits. 


\begin{table}[h]
\caption{Comparison of methods using Binary Encoding}
\label{tab:resource}
\centering
\begin{tabular}{lcccccc}
\toprule
\multicolumn{2}{c}{\textbf{Grover Full Quant}}&\multicolumn{2}{c}{\textbf{Grover Quant-Clas}}&\multicolumn{2}{c}{\textbf{QAOA}}\\
\multicolumn{2}{c}{\textbf{\# Qubits=25}}&\multicolumn{2}{c}{\textbf{\# Qubits=14}}&\multicolumn{2}{c}{\textbf{\# Qubits=8}}\\
\midrule
\textbf{\# Iters} & \textbf{\% valid} & \textbf{\# Iters} & \textbf{\% valid} & \textbf{\# layers} & \textbf{\% valid}\\
\midrule
1 &  52.5& 5 & 53.6 & 5 & 31.5\\
2  & 94.8& 10 & 60.3 & 10 & 36.7\\
3& 90.3&15&51.2&15& 55.3\\
\bottomrule
\end{tabular}

\end{table}
\subsection{Analysis of Number of Iterations}
Ideally in Grover algorithm after sufficient iterations, $\eta_v$ should approach 1, i.e., probability mass is concentrated on valid solutions. However in experiments, $\eta_v$ first increased as Grover amplifies valid states, then decreased after the optimal iterations is exceeded as in Table~\ref{tab:resource}, due to oscillatory behavior of amplitude amplification. While for \emph{QA} the anneal time $T_a$ was a deciding factor, with increasing $T_a$ $\eta_v$ increased. 

\subsection{Run-Time Analysis}
Experimentally, the order of runtime is found to be $R_{\text{grover}}>R_{\text{QAOA}}>R_{\text{QA}}$, though this ordering can shift with increasing Grover iterations, QAOA layers, or anneal time. For the reduced graph $G_q$ in Fig.~\ref{fig:method}, Grover iterations $1,2,3,4$ take $65.0, 119.5, 177.4, 238.7$ seconds respectively, while QAOA with $5, 10, 15, 25$ layers takes $13.5, 21.5, 31.8, 52.9$ seconds respectively. Quantum annealing over $150, 200, 250$ anneal steps takes only $4.8, 6.1, 8.2$ seconds respectively, but with a markedly lower valid fraction than Grover or QAOA at comparable resource budgets, consistent with QA's lack of an explicit constraint-oracle mechanism (Table~\ref{tab:compare}).

\subsection{Solver and Encoding Selection}
Algorithm~\ref{alg:solver} illuminates the path for selecting solvers based on graph type and resource constraints, consistent with the suitability column of Table~\ref{tab:compare}. Algorithm~\ref{alg:constraint} presents the corresponding choice of encoding depending on the solver and the qubits available.

\subsection{Reduction Analysis}
Using the reduction techniques of Section~\ref{sec:Red_methods}, qubit count, gate count, and runtime can all be significantly reduced. For example, $Q_{\text{logical}}$ under binary encoding for the graphs in Fig.~\ref{fig:method}(a), (b), and (c) reduces from 12 to 8, 12 to 6, and 24 to 14, respectively, after applying the corresponding reduction (Results 1--4, Section~\ref{sec:Red_methods}). The general reduction benefit for any method in Table~\ref{tab:compare} can be estimated by substituting $N$ with the post-reduction vertex count $N_{\text{red}}=|V_q|$ throughout.

\subsection{Limitations of the Current Evaluation}
The runtime and valid-fraction figures reported above come from a single small benchmark graph ($N=4$, $K=3$; Fig.~\ref{fig:method}) executed on a classical simulator, and are intended to illustrate qualitative trends -- e.g., the amplification behavior of Grover iterations and the depth/qubit trade-offs across solvers -- rather than to establish statistically robust performance claims. We have not yet compared against classical baselines (e.g., greedy or DSATUR coloring) on the same instances, nor evaluated on noisy or hardware backends, both of which are necessary before drawing conclusions about practical quantum advantage; we leave a broader empirical study across graph families, sizes, and noise models to future work.

\section{Conclusion}
\label{sec:conclusion}
The paper presented a resource-efficient solver-agnostic quantum framework, for $K$-coloring, emphasizing binary encoding and optimized constraint circuits, that can optionally incorporate classical symmetry reduction as a pre-processing step. The approach significantly reduces qubit count compared to one-hot encoding for \emph{QAOA} and $Grover_{Q-Cls}$ enabling the simulation or execution of larger instances on near-term quantum hardware. Optimal encoding and constraints specific to for Grover search, \emph{QAOA}, and quantum annealing based K-coloring was presented.

\vspace{12pt}

\newpage
\appendices
\end{document}